# Fabricar os dados: o trabalho por trás da Inteligência Artificial

Matheus Viana Braz[1]
Paola Tubaro[2]
Antonio A. Casilli[3]

**Introdução**

A produção da Inteligência Artificial (IA) é comumente associada ao trabalho de engenheiros de software ou profissionais altamente qualificados, vinculados à grandes empresas ou start-ups especializadas, que se desenvolveram inspiradas na ideologia californiana (Brook, Cameron, 2015). É pouco difundido, contudo, que o desenvolvimento de "tecnologias inteligentes" depende, em diferentes etapas, de uma multidão de trabalhadores precarizados, sub-remunerados e invisibilizados, os quais dispersos globalmente realizam atividades repetitivas, fragmentadas, pagas por tarefa e feitas em poucos segundos.

Tratam-se de trabalhadores que rotulam dados para treinarem algorítmos, mediante tarefas que necessitam das capacidades intuitivas, criativas e cognitivas dos seres humanos, tais como categorização de imagens, classificação de publicidades, transcrição de áudios e vídeos, avaliação de anúncios, moderação de conteúdos em mídias sociais, rotulagem de pontos de interesse anatômicos humanos, digitalização de documentos etc. (Gray, Suri 2019; Casilli 2019). Embora se trate de um processo essencial ao aprendizado de máquinas (*machine* ou *deep learning*), esse trabalho é externalizado para plataformas digitais ou para redes especializadas de terceirização (ILO 2021), bem como é realizado nas franjas da informalidade, sem quaisquer proteções social ou trabalhista (salvo algumas exceções), tampouco autonomia para negociação de remuneração.

Esta forma de trabalho é frequentemente designada por "microtrabalho". Este termo ganhou popularidade em meados dos anos 2000, baseado no conceito de "microcrédito", abordagem econômica que teria supostamente por finalidade garantir inclusão financeira à populações vulnerabilizadas e marginalizadas, sem acesso a serviços bancários. É nessa esteira discursiva que surgem as plataformas digitais especializadas em rotulagem e treinamento de dados para IA, como se a produção de dados pudesse oferecer uma possibilidade semelhante, porém relacionada a obtenção de renda online. No mundo todo, estima-se que haja mais de 160 milhões de trabalhadores registrados em plataformas de microtrabalho e *freelancer* (Kässi, Lehdonvirta, Stephany 2021). Nas plataformas de trabalho *freelancer,* a média global de remuneração é de US$7.6, enquanto nas de treinamento de dados a média é de US$3.3 (ILO, 2021).

---

[1] Professor Adjunto do Departamento de Psicologia da Universidade do Estado de Minas Gerais (UEMG), Brasil, e Professor do Programa de Pós-graduação em Psicologia da Universidade Estadual de Maringá (UEM). E-mail: matheus.braz@uemg.br

[2] Professora pesquisadora (diretora de pesquisa) no Centro Nacional de Investigação Científica (CNRS). Socióloga especializada em ciência da computação, vinculada ao Centro De Pesquisa em Economia e Estatística (CREST). E-mail:paola.tubaro@cnrs.fr

[3] Professor de Sociologia no Instituto Politécnico de Paris -Telecom Paris. Codiretor do DiPLab (Digital Platform Labor) e cofundador da Rede Internacional de Trabalho Digital (INDL). E-mail: antonio.casilli@ip-paris.fr

Os estudos sobre o microtrabalho surgiram em meados de 2010 (Ipeirotis 2010; Irani, 2015) e a maior parte se concentra em países da Europa e América do Norte. Os principais produtores de IA no mundo estão sediados no Norte Global, porém se servem majoritariamente de mão de obra barata proveniente do Sul Global (ILO 2021). Pesquisas recentes evidenciaram ainda a expressiva presença de trabalhadores de diferentes países da América Latina nessa cadeia produtiva (Miceli et al. 2020; Miceli, Posada 2022; Posada 2022; Schmidt 2022).

No Brasil, Grohmann (2021) e Antunes (2020; 2023) organizaram obras coletivas dedicadas à compreensão dos impactos das plataformas digitais no mundo do trabalho, inclusive no que diz respeito à produção da IA e a nova morfologia do trabalho em nossa sociedade. A plataformização do trabalho e a uberização, nesse sentido, tem sido objeto de inúmeros debates, notadamente relacionados ao processo histórico de informalização do trabalho em território nacional (Abílio 2017; 2019; Grohmann 2020; Abílio, Amorim, Grohmann 2021; Manzano, Krein, Abílio 2023). Todavia, ainda são incipientes as pesquisas empíricas concernentes ao trabalho digital de rotulagem e treinamentos de dados para IA (Kalil 2019; Moreschi et al. 2020; Viana Braz 2021; 2022; Grohmann, Araújo 2021).

Em contraposição, Viana Braz (2021) constatou a existência de ao menos 54 plataformas de microtrabalho em operação no país, sendo que 29 abarcavam microtarefas de geração, anotação ou verificação de dados para aprendizagem de máquinas, o que indica que o Brasil constitui uma das principais reservas de mão de obra desse mercado na América Latina. Mas quem são, afinal, os trabalhadores brasileiros que atuam nessas plataformas? Em quais condições realizam seus trabalhos? Quanto ganham? Por que recorrem às plataformas? Quais as diferenças entre trabalhadores do Norte e Sul Global, quando nos remetemos à produção e qualificação de dados que sustentam a produção da IA? Esse trabalho poderia, enfim, ser considerado como uma verdadeira oportunidade de renda para países de subdesenvolvidos ou em desenvolvimento?

Para responder à essas interrogações, buscamos cartografar o perfil sociodemográfico dos trabalhadores brasileiros no mercado de anotação de dados, para depois analisarmos em quais condições realizam suas atividades. Para tanto, realizamos uma pesquisa exploratória que abarcou métodos mistos. Primeiro, realizamos uma etnografia digital entre os anos de 2020 e 2021, em 24 grupos de WhatsApp e Facebook relacionados a esse mercado (Viana Braz, 2021). A partir dessa inserção no campo, no ano seguinte convidamos individualmente alguns membros desses grupos e realizamos 15 entrevistas em profundidade. Tratou-se de 10 mulheres e 05 homens, com idade mínima de 22 e máxima de 54 anos. Entre os participantes, 13 pessoas possuíam ensino superior completo, em diferentes áreas, como direito, engenharia de petróleo, fisioterapia, engenharia civil, administração, biotecnologia, letras, comércio exterior e ciência da computação. Todos trabalhavam em ao menos uma e no máximo quatro plataformas concomitantemente, entre as seguintes: Amazon Mechanical Turk, Appen, Telus (anteriormente denominada Lionbridge), Clickworker, Quadrant e OneForma.

Em 2023, nos servimos de metodologia análoga àquela utilizada nos trabalhos de Casilli et al. (2019), Kalil (2020), Moreschi (2021) e aplicamos um questionário online à 477 trabalhadores brasileiros que atuavam na plataforma Microworkers. O instrumento foi desenvolvido no bojo do DiPLab (Digital Platform Labor research group), traduzido da versão em espanhol (aplicado em países como Argentina, Bolívia, Chile, Colômbia, Costa Rica, Equador, República Dominicana, Venezuela, etc) e adaptado para o português, de acordo com particularidades do Brasil. O questionário compreendia cerca de 120 questões, relacionadas à informações sociodemográficas variadas, como níveis de escolaridade, situação familiar, experiência profissional, renda, uso da internet, relações sociais e condições de trabalho nas plataformas. A escolha pela plataforma se justifica pelo fato de abarcar variadas modalidades de microtrabalho. No que concerne às questões éticas específicas das pesquisas em plataformas de microtrabalho, nos fundamentamos nas reflexões e recomendações realizadas por Molina et al. (2023). Para análise dos dados, primeiro fizemos uma caracterização das plataformas de treinamentos de dados, de maneira a compreendermos como elas se inserem em cadeias globais de suprimentos para produção da IA. Depois, problematizamos a sociodemografia e condições de trabalho dos brasileiros nesse mercado, com foco nas assimetrias existentes entre países do Norte e Sul Global.

**Fabricando os dados e invisibilizando o trabalho**

Pedro, 54 anos, é graduado em comércio exterior, porém trabalha como corretor. Concilia essa atividade com outro trabalho, em plataformas digitais cujas tarefas implicam atividades como transcrever áudios, classificar avaliações de anúncios em websites de empresas e catalogar descritores para aperfeiçoar os mecanismos de busca do Google. No passado, participou também de projetos relacionados à moderação de conteúdos pornográficos e violentos em mídias sociais como Facebook e Instagram. Clara, 38 anos, é graduada em administração e morou no exterior por alguns anos. Ao voltar para o Brasil, trabalhou como professora de inglês, mas não se adaptou à escola à qual estava vinculada. Desde então, trabalha em plataformas de microtarefas variadas, notadamente relacionadas à avaliações de anúncios, classificação de dados e moderação de conteúdos em mídias sociais. Izabela, por sua vez, tem 29 anos e é mãe solo de uma filha de 09 anos. Mudou-se do nordeste para o sudeste do país para fazer um mestrado em uma prestigiada universidade pública. Como sua bolsa de pesquisa não lhe oferece condições financeiras suficientes para se manter, trabalha todas as noites em plataformas como Appen e Oneforma, realizando microtarefas de transcrições de áudios. Embora possuam trajetórias de vida bastante distintas, o que Pedro, Clara e Izabela compartilham em comum?

Ambos trabalham em plataformas digitais, como Appen, Oneforma, Telus, Clickworker e Microworkers, voltadas à rotulagem de dados para o desenvolvimento da Inteligência Artificial (IA). Se os dados são uma das principais fontes de valor na indústria de "tecnologias inteligentes" (como carros autônomos, /assistentes de voz e chatbots como o ChatGPT), fabricá-los e treiná-los é mais complexo do que comumente imaginamos. De acordo com Tubaro et al. (2020), o trabalho de geração e anotação de dados é vastamente utilizado na produção da IA e cumpre uma variedade de funções que gravitam em torno de três pólos: *preparação, verificação* e *imitação da IA*, conforme evidenciado na Figura 1:

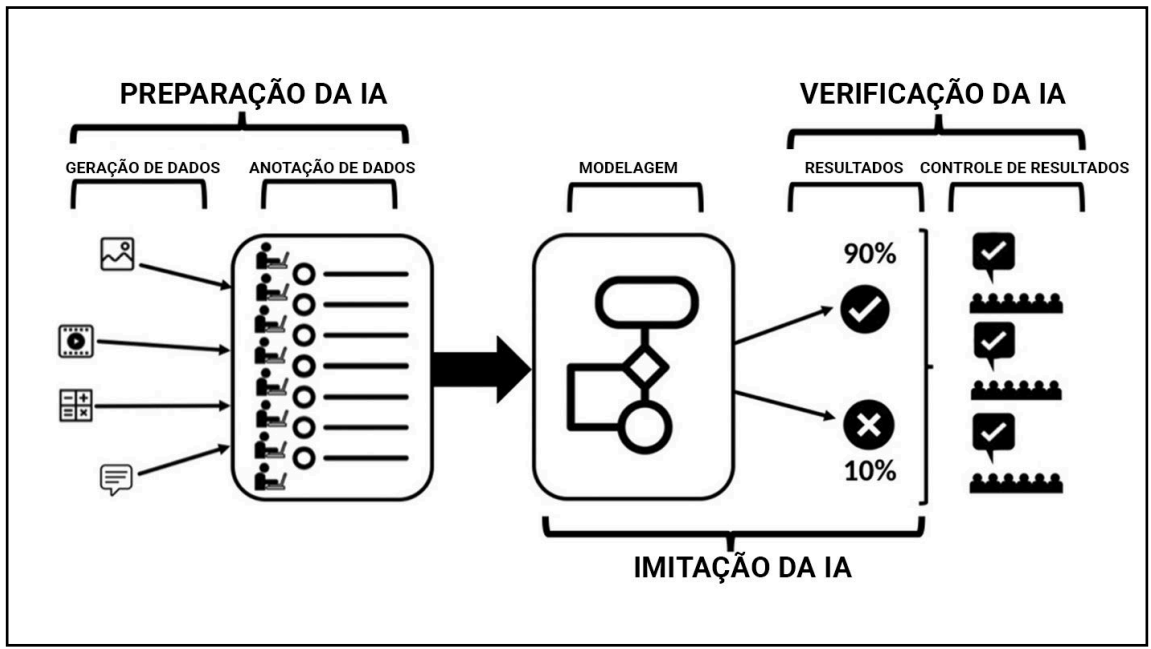

Figura 1: *Três principais funções do microtrabalho no desenvolvimento de soluções de IA baseadas em aprendizado de máquina e uso intensivo de dados.* Traduzido de Tubaro et al (2020).

No primeiro caso, para que empresas, start-ups ou laboratórios de pesquisa desenvolvam algum tipo de IA, extensas bases de dados precisam ser geradas e qualificadas (via anotação e rotulagem), para que sejam estabelecidos os parâmetros técnicos dos algoritmos de aprendizagem. Esse processo exige um trabalho cognitivo sensível, elementar para extração e geração de valor de tais dados (Viana Braz, 2021). Na indústria de reconhecimento facial, por exemplo, a preparação da IA compete a trabalhadores que, mediante plataformas digitais ou redes especializadas de terceirização, realizam manualmente microtarefas de geração de dados, identificação de padrões de reconhecimento, classificação de expressões faciais e rotulagem de diferentes partes anatômicas do rosto (nariz, boca, olhos etc) (Tubaro et al. 2020).

Mesmo após os processos de geração e anotação de dados, os parâmetros técnicos da produção da IA precisam ser continuamente aperfeiçoados, sobretudo para correção de eventuais falhas e para garantir maior acurácia técnica dos resultados dos algoritmos de aprendizagem. E aqui nos remetemos à função de verificação dos resultados da IA. Em plataformas digitais como Appen e Clickworker, microtarefas de verificação são recorrentes e envolvem, por exemplo, ouvir áudios e verificar se a transcrição automática gerada pelo assistente virtual está correta ou não (Tubaro et al. 2020). Nesse cenário, Perrigo (2023) revelou em reportagem publicada na revista *Time*, que visando deixar o ChatGPT menos "tóxico", a OpenAI (empresa proprietária) contou com a terceirização de quenianos, contratados por menos de dois dólares por hora, para realizarem tarefas de verificação e rotulagem de dados relacionadas à comentários violentos que descreviam detalhadamente situações de abuso sexual infantil, assassinato, incesto, zoofilia, suicídio, tortura e automutilação. A referida empresa, denominada Sama, é sediada em São Francisco e se especializou na terceirização de trabalhadores no Quênia, Uganda e Índia, para rotular dados para clientes como Microsoft, Google e Meta.

Em outras situações, como o desenvolvimento do aprendizado de máquinas é caro e por vezes falho, trabalhadores passam a desempenhar funções que supostamente deveriam ser realizadas por máquinas. Trata-se da imitação da Inteligência Artificial. É o caso, por exemplo, do Google Duplex, assistente virtual voltado à realização de ligações à empresas para agendamento de reservas e confirmação de horários. Em 2019, contudo, descobriu-se que cerca de 25% das chamadas realizadas pela ferramenta eram feitas por humanos em *call centers*, os quais imitavam a IA (Chen, Metz 2019).

Nessa mesma direção, Le Ludec et al. (2023) relatam o caso de uma start-up francesa, que criou uma IA centrada em visão computacional a qual permite que câmeras de vigilância em lojas identifiquem furtos de maneira automatizada. Ao perceber algum tipo de comportamento suspeito, os caixas ou gerentes de lojas recebem mensagens automáticas de alerta em seus celulares. Três empresas terceirizadas em Madagascar e uma na Indonésia fazem o microtrabalho de preparação de dados e verificação dos resultados da IA da companhia na França, que envolve majoritariamente assistir a vídeos de câmeras (com pessoas supostamente roubando, desembalando ou danificando produtos) e rotular comportamentos considerados suspeitos, em menos de um minuto. Os autores revelaram, todavia, que alguns vídeos à serem rotulados eram, na realidade, reproduzidos em tempo real, diretamente das câmeras de alguns clientes da start-up. Mais do que preparação e verificação, trabalhadores imitavam a IA e cumpriam a função de seguranças remotamente (Le Ludec et al. 2023).

Processamentos de linguagem natural (que utilisam *Large Language Models),* visão computacional, análise preditivas, sistemas autônomos, assim como toda tecnologia baseada em aprendizagem de máquina, incluindo a Inteligência Artificial Generativa, dependem vastamente do trabalho humano de preparação, verificação ou imitação da IA. À título de ilustração, segundo relatório produzido pela Cognilytica (2020), estima-se que o mercado de geração e anotação de dados para IA chegará a 3,5 bilhões de dólares até o fim de 2024. Tarefas de preparação e rotulagem de dados, nesse contexto, representam cerca de 80% do tempo gasto na maior parte dos projetos de IA que envolvem aprendizagem de máquina (Cognilytica 2020). Evidencia-se, portanto, que esse trabalho constitui um ingrediente secreto da cadeia de suprimentos da IA.

Secreto, pois ele parece ser invisibilizado, por parte dos produtores da IA globalmente, de modo que sobretudo no Brasil, segue ainda distante da esfera pública e da agenda regulatória, no âmbito na IA como no do trabalho plataformizado.

Nos últimos 10 anos, esse mercado se diversificou e se expandiu de maneira heterogênea. A Amazon Mechanical Turk, primeira plataforma global de treinamentos de dados, lançada em 2005 por Jeff Bezos, parece ter perdido relevância, frente ao surgimento de novas plataformas intermediárias (ILO 2021) e a criação do Sagemaker Ground Truth, uma plataforma interna que usa tarefas humanas para aprimorar os modelos de aprendizado de máquina. Schmidt (2022), nessa direção, discorre que a demanda de maior acurácia e complexidade das tarefas resultou num processo de especialização das plataformas em mercados e projetos determinados.

Existem plataformas de anotações de dados focadas em microtarefas de uso geral (*general-purpose crowdwork*), na qual as tarefas são ofertadas numa espécie de *market place* aberto, como ocorre na Neevo, Microworkers, Hive Work e Amazon Mechanical Turk (Schmidt 2022). Há, também, plataformas construídas por meio de múltiplas camadas e níveis de complexidade (*specialized full-service crowd-AI stacks*), como Appen, Telus, Quadrant. Mighty AI e Scale IA. São organizadas mediante projetos específicos, nos quais os trabalhadores precisam fazer provas, para atestar habilidades e conhecimentos pré-definidos. Ao serem aprovados, após assinarem um acordo de confidencialidade (*non-disclosure agreement*), estabelece-se um contrato de tempo parcial e são direcionados para outra camada nas plataformas. Nela, deve-se cumprir uma quantidade específica de horas ou tarefas por dia e a remuneração ocorre mediante um valor/hora definido pela plataforma (Schmidt 2022).

No mundo todo, os trabalhadores dedicam em média 18.6 horas semanais à atividades remuneradas nessas plataformas e 32% deles tem como principal fonte de renda o trabalho de anotação de dados (Berg et al. 2018). Como o processo de distribuição e avaliação das atividades é controlado por algorítmos, os preços são definidos unilateralmente, assim como é comum que hajam rejeições injustas das tarefas realizadas, de modo que os trabalhadores não recebem por trabalhos concluídos e não raro tampouco recebem justificativas dos motivos da rejeição, pois não há comunicação entre o cliente e o trabalhador. Além de culminar na perda de rendimentos, as reprovações reduzem suas classificações nas plataformas, o que limita suas possibilidades de acederem a melhores projetos (ILO 2021).

Ainda que os modelos de negócios tenham se diversificado nos últimos anos, as diferenças de remuneração e condições de trabalho desses mercados evidenciam a criação de canais mundiais de deslocamento da força de trabalho, no qual atividades mal pagas e invisibilizadas refletem assimetrias e tensões históricas em relação ao Norte e Sul Global. Em países desenvolvidos, a média de rendimentos é US$4.00, contra US$2.1 em países em desenvolvimento. Ademais, em estudo conduzido pela Organização Internacional do Trabalho (Berg et al. 2018), que abarcou cinco plataformas em 75 países, constatou-se que o rendimento médio por hora dos trabalhadores nessas plataformas é de US$4.70 na América do Norte, US$3.00 na Europa e Ásia Central, US$2.22 na Ásia Pacífico e US$1.33 na África. Descortina-se, nesse contexto, uma série de desigualdades digitais (Robinson et al. 2020a, 2020b), intensificadas pela pandemia da COVID-19 (Robinson et al. 2020c) e relacionadas à marcadores de raça (Van Doorn 2017), gênero (Morell 2022; Tubaro et al. 2022), classe, educação (Tubaro 2022) e território (Braesemann et al. 2022).

No Brasil, Viana Braz (2021) observou que o trabalho de anotação e rotulagem de dados se insere num ecossistema mais amplo, de plataformas de microtrabalho que se difundem por meio da promessa de renda-extra online, de maneira supostamente fácil e rápida. Esse mercado abarca também plataformas de fazendas de cliques (*click farms*), voltadas ao mercado de compra e venda de seguidores, curtidas, comentários e inscritos em mídias sociais como Instagram, Facebook, Youtube, TikTok, Kwai e Spotfy (Viana Braz, 2021; Grohmann et al. 2022). Carece na literatura, no entanto, uma caracterização desses trabalhadores, para compreender as nuances e particularidades presentes em território nacional.

**Quem são os trabalhadores brasileiros por trás da Inteligência Artificial?**

Os trabalhadores brasileiros são em sua maioria jovens, com idade entre 18 e 35 anos (70,6%), mulheres (63,9%) e casados, vivem com parceiros ou possuem união estável (60,8%). Os três estados brasileiros com maior presença de trabalhadores foram São Paulo (28,8%), Rio de janeiro (12,6%) e Minas Gerais (9,7%). As taxas de escolarização observadas foram maiores que as médias da população brasileira. Chama atenção que enquanto cerca de 20% da população brasileira acima de 25 anos possui ao menos Ensino Superior completo (IBGE, 2022), esse montante é de 49,5% na referida amostra, níveis estes correspondentes à média (47,4%) encontrada entre os países da Organização para a Cooperação e Desenvolvimento Econômico (OCDE, 2023).

Tal como constatado nos estudos de Moreschi (2020) e Grohmann e Araújo (2021), parece ser comum a utilização de múltiplas plataformas, inclusive concomitantemente (Viana Braz, 2021). Parte significativa dos trabalhadores da Microworkers (n=342/477) declarou já ter trabalhado em outras plataformas. Dentre as mais utilizadas, destacam-se, Clickworker (48,6%), Appen (37,5%) Oneforma (28,1%), Picoworkers (24,5%), UHRS (14,9%) e Telus (13,1%). Além disso, mais de 50% dos trabalhadores relataram já ter realizado atividades relacionadas à vendas online, jogos de azar ou apostas esportivas online, o que confirma que o trabalho de anotação de dados para IA está inserido em um ecossistema amplo, direcionado à obtenção de renda extra na internet. Diante do aumento da precarização do trabalho, em um país no qual cerca de 40% da força de trabalho (38 milhões de pessoas) está na informalidade (IBGE, 2023), o trabalho plataformizado parece ter emergido como mais uma alternativa de renda, sobretudo para jovens, substancialmente qualificados, porém que não encontram no mercado formal de empregos condições suficientes para se sustentarem.

Necessidade de dinheiro, flexibilidade de horários e preferência por trabalhar em casa são as principais motivações que levam os trabalhadores a realizarem microtarefas. Em consonância com o estudo de Berg et al. (2018), 33,5% dos participantes tem como única fonte de renda as plataformas, porém os rendimentos constatados foram expressivamente inferiores. Em média, os brasileiros recebem US$1.80 por hora (valor análogo aos pagamentos feitos aos quenianos, subcontratados pela OpenAI), o que corresponde a menos de dez reais por hora. Tratam-se de trabalhadores que dedicam cerca de 15 horas e 30 minutos semanais à realização de microtarefas, para receberem, em contrapartida, em média R$582,71 reais mensais.

Contando com todas suas fontes de renda (incluindo a rotulagem de dados), R$1866.00 é o rendimento médio mensal dos trabalhadores, o equivalente a menos de 1,5 salário mínimo em 2023. Em cidades como Rio de Janeiro, São Paulo e Belo Horizonte, segundo o Departamento Intersindical de Estatística e Estudos Socioeconômicos (DIEESE, 2022), o custo mensal estimado para uma única pessoa é aproximadamente de R$5400,00.

Ao nos aprofundarmos nos dados coletados, as características da informalização do trabalho (Manzano et al. 2023) nessa população ficam ainda mais evidentes. Por exemplo, 66% dos trabalhadores contam com uma quantidade mínima de dinheiro a ser obtida nas plataformas para o pagamento de suas contas. Mais ainda, 49,8% tiveram pelo menos dois trabalhos informais e 68,9% ao menos três empregos formais ao longo da vida. Entre os participantes que estavam contratados formalmente, 40,5% trabalhavam em tempo parcial, enquanto a média global é de 33% (Berg et al. 2018). Dentre eles, 72% atuavam em ocupações que exigiam alta qualificação. Este valor se distingue das taxas de 65% nos países da América Latina e Caribe, 61% na Ásia e no Pacífico, 59% na Europa e Ásia Central e menos de 20% na América do Norte (Berg et al. 2018).

Contrasta-se com a literatura (Berg et al. 2018; Kalil, 2019; Moreschi et al. 2020, ILO, 2021) o fato que a maior parte dos participantes (63,9%) são mulheres, o que parece ser uma particularidade do Brasil em relação a outros países, onde comumente a maioria dos trabalhadores são homens. Contudo, em consonância com os estudos feitos pela Organização Internacional do Trabalho (ILO, 2021), os rendimentos das mulheres são ligeiramente superiores aos dos homens, em parte porque elas entram nas plataformas com mais frequência e realizam microtarefas em horários mais bem remunerados. À título de elucidação, 67,9% das mulheres e 55,8% dos homens fazem login pelo menos uma vez por dia para procurar novas tarefas remuneradas nas plataformas. Enquanto a maior proporção de homens (43,6%) trabalha nestas plataformas fora do horário comercial (das 18h00 às 22h00), 54,8% das mulheres trabalham habitualmente entre as 14h00 e as 18h00. Nossos resultados confirmam os principais achados do estudo de Tubaro et al. (2022), que sugere que as mulheres tendem a entrar nas plataformas por períodos mais curtos e com maior frequência do que os homens. Isso indica que o tempo livre dessas mulheres parece ser cada vez mais direcionado para a realização de microtarefas.

Ainda no âmbito das assimetrias de gênero, entre os participantes desempregados, 73,7% são mulheres. Além disso, 38,7% das mulheres dependem exclusivamente de plataformas para obter rendimentos, em comparação com 24,1% dos homens. Enquanto 55,2% dos homens são assalariados, somente 41,3% das mulheres possuem empregos formais. No que diz respeito ao tempo dedicado a tarefas domésticas (como fazer compras, limpar, cuidar dos filhos, cozinhar etc.), as mulheres e os homens relatam gastar em média 13 horas e 48 minutos e 8 horas e 37 minutos por semana, o que corresponde a uma diferença de 37,5%.

Além disso, 62,6% das mulheres são mães ou tutoras legais de um ou mais filhos, em comparação com 45,3% dos homens. Em termos de apoio financeiro, 62,3% das mulheres e 39,5% dos homens dependem dos seus parceiros para sustentar as suas famílias. Por fim, 53,5% dos homens são os principais provedores de seus lares, enquanto para as mulheres esta porcentagem cai praticamente para a metade (26,9%). Em resumo, constatamos que as plataformas não oferecem uma alternativa sólida e estável de rentabilidade aos trabalhadores. Portanto, encerra-se as mulheres em um círculo insidioso, pois o trabalho de rotulagem de dados se soma a outros trabalhos invisíveis feitos dentro de suas casas, ao mesmo tempo que não garante a elas autonomia financeira.

Observamos também que o trabalho de rotulagem de dados se opera mediante uma estratégia de dispersão por parte das plataformas (Viana Braz, 2022), que não permitem que os trabalhadores entrem em contato uns com os outros. Com efeito, 69,6% não conhecem mais ninguém que trabalhe nas plataformas. Apenas 22,2% da amostra relatou participar em fóruns ou comunidades de discussão online sobre o microtrabalho. Entre elas, as plataformas mais populares são WhatsApp e Telegram (53,2% e 15,3%, respectivamente). Dentre esses trabalhadores, 45,4% alegam que fazem parte desses grupos para conversar com outras pessoas que trabalham online. Compartilhar informações sobre tarefas criticadas por outros trabalhadores, queixar-se das plataformas e se atualizar sobre tarefas disponíveis são outras motivações para participar de tais espaços.

É revelador, nesse cenário, que as principais queixas dos trabalhadores abrangem a instabilidade financeira, a falta de transparência algorítmica e, em especial, o cansaço e a falta de interação entre eles (Viana Braz, 2022). Mencionaram, ademais, que as "piores tarefas" seriam aquelas relacionadas à moderação de conteúdos violentos e pornográficos nas mídias sociais. Há também aquelas caracterizadas como "estranhas", que também entrariam nesse rol. Helena, 54 anos, por exemplo, nos relatou que trabalhara em um projeto voltado à geração de dados de "robôs aspiradores de pó", para que o software identificasse e evitasse passar por cima de fezes de cachorros. As microtarefas consistiam em tirar "fotos de cocôs de tais animais" em variados ambientes domésticos. Alguns centavos de dólares eram pagos para cada foto enviada. A trabalhadora nos relatou que passara dois dias movendo as fezes de seu cachorro e chegou a tirar mais de 250 fotos em diferentes locais de sua residência.

Quando se trata de moderar conteúdos violentos e pornográficos nas redes sociais, os trabalhadores manifestaram preocupações com o custo psicológico de tais atividades e com a falta de apoio que recebem das plataformas. Revolta, incômodo, impotência e tristeza foram alguns dos sentimentos relatados pelos trabalhadores nessas tarefas (Viana Braz, 2022), conforme ilustrado nos relatos de Lucas e Pedro:

> Eu trabalhava em um projeto do Facebook, tinha que verificar o anúncio, pra avaliar se tinha sangue, violência, abuso, se continha arma. Muitas vezes, peguei anúncio pesado. [...] Você precisa ter um psicológico forte pra trabalhar nisso. Você tinha que fazer tudo dentro de uma hora. Eles só te pagam o valor referente à hora. Eles falam que se você for ver e não conseguir terminar nem começa. Precisava de ter um apoio psicológico, um amparo. Uma mulher que conheci teve que fazer tratamento. É preciso tentar minimizar o impacto dos trabalhadores que ficam assistindo a mortes, pra tirar a imagem da cabeça da pessoa, porque ninguém consegue acostumar com isso (Lucas, 23 anos).

> Não dá para se sentir bem com isso, né? A gente sente nojo do ser humano, infelizmente! Pensar que esse tipo de ação vem de uma espécie como a sua, é horrível você ver como o ser humano é capaz de fazer esse tipo de coisa (Pedro, 54 anos).

Embora essas tarefas suscitem angústia e sentimentos ansiogênicos aos trabalhadores, as plataformas parecem se eximir dos riscos psicossociais concernentes a tais atividades. Conforme destacado em estudo anterior (Viana Braz, 2022), no acordo de confidencialidade da OneForma, por exemplo, precisamente no item 5, nomeado *Isenção de conteúdo adulto* (adult contents waiver), consta o seguinte: "O contratado está ciente da possível existência de conteúdos adultos em materiais transmitidos como parte dos trabalhos nos projetos através do website e, consequentemente, aceita a mencionada possibilidade, 'renunciando a todas as reclamações decorrentes desse fato'"[4] (tradução nossa). Igualmente, no acordo de participação da Amazon Mechanical Turk, precisamente no item 2, a plataforma é enfática ao destacar que "não somos responsáveis pelas ações de nenhum Requisitante ou Trabalhador [...]. Seu uso do site ocorre por sua conta e risco"[5] (Viana Braz, 2022).

Roberts (2019) chama atenção para o fato de que não há estudos longitudinais suficientes na literatura acerca dos efeitos do trabalho de moderação de conteúdos na saúde mental dos trabalhadores. Nosso estudo não oferece dados conclusivos a esse respeito, contudo constatamos uma falta de amparo psicológico e espaços de apoio para que esses trabalhadores expressem seus sofrimentos. Na prática, as estratégias

---

[4] No original: "Contractor is aware of the possible existence of adult contents in materials transmitted as part of work projects through the web site and, as a result of this, accepts the mentioned possibility, waiving all claims arising from this fact".

[5] No original: "We are not responsible for the actions of any Requester or Worker […]. Your use of the site is at your own risk".

encontradas para preservação de saúde são notadamente individualizadas (Viana Braz, 2022).

Os processos de gerenciamento e controle algorítmico sobre o trabalho tampouco são claros. Os trabalhadores se queixam sobre a nebulosidade das políticas de desligamentos (e bloqueio) e dos critérios de admissão em projetos, assim como de aprovação e rejeição das tarefas nas plataformas. Claudia, 38 anos, por exemplo, que dedicava horas do seu dia à rotulagem de dados, nos contou sobre sua experiência ao ser desligada repentinamente de um projeto em uma plataforma:

> Depois que eu fui demitida [da plataforma], tive uma crise bem horrível, aí tive que procurar o psiquiatra e ele me medicou. [...] É revoltante, dói bastante, de repente você fica sem o seu trabalho e em meio a uma pandemia. Eu mandei vários e-mails quando fui demitida, mas não recebi nenhuma resposta. [...] Teve uma época que estava sem task para fazer, a não ser que você acordasse 3 ou 4 horas da manhã para fazer. Tenho uma conhecida que fez isso e ela ficou muito doente. Mas, assim mesmo, tem muita gente que faz, agora vai entrar abril e maio, quem acordar às 3 é que vai conseguir trabalhar (Cláudia, 38 anos).

Nessa direção, as ofertas de tarefas nas plataformas são imprevisíveis e frequentemente não são claras as formas como são distribuídas. Como a maior parte das empresas são sediadas no Norte Global, os trabalhadores queixam-se que a reposição de tarefas costuma ocorrer de acordo com os fusos-horários dos clientes-sede, o que coloca os brasileiros em desvantagem na realização de tarefas em projetos globais. Uma das alternativas para contornar essa desvantagem é passar a trabalhar no período da madrugada nos momentos onde ocorrem a reposição. Isso explica, provavelmente, porque 27,9% dos participantes trabalham nas plataformas entre 22h e 01 da manhã e 9,4% entre 01h e 05h da manhã.

**Considerações finais**

A IA gera tanto entusiasmo quanto desilusão, com promessas que muitas vezes não são cumpridas. Por isso, não é de surpreender que o trabalho humano, que é seu componente fundamental, também esteja sujeito a essas mesmas decepções. Na prática, para os trabalhadores, as promessas de desenvolvimento econômico e social que o conceito de microtrabalho inicialmente oferecia não se materializaram. Não só os salários são baixos, mas, acima de tudo, as condições de trabalho são duras. A disponibilidade altamente variável de tarefas leva à migração de uma plataforma para outra, resultando em tempo de trabalho não remunerado. A necessidade de sincronização com clientes estrangeiros força os microtrabalhadores a ficarem na frente de seus computadores durante a noite. Além disso, trabalhar em casa, sem encontrar colegas ou clientes, os isola e, como vimos, a moderação de conteúdo exigiria apoio psicológico e médico, o que em grande parte não existe. É essencialmente porque outras pessoas não estariam preparadas para trabalhar nessas condições que essa atividade é deixada para grupos relativamente desfavorecidos: mães, desempregados, trabalhadores com contratos precários ou que trabalham na economia informal. Seu acesso residual ao microtrabalho, sem medidas de apoio especificamente voltadas para eles, os coloca na invisibilidade.

Contudo, essas pessoas têm habilidades linguísticas, culturais e de informática, que mobilizam para o microtrabalho. Elas precisam saber como pesquisar, classificar e avaliar as informações on-line; comunicar-se em inglês ou, pelo menos, dominar os plug-ins de tradução automática; comparar tarefas e plataformas para selecionar a mais lucrativa; escolher um método de pagamento calculando os encargos bancários e as tendências das taxas de câmbio. Há também um processo de aprendizado ao longo do tempo, que leva a um trabalho mais rápido, ampliando a gama de tarefas que os microtrabalhadores podem executar e, em alguns casos, até mesmo automatizando às atividades por meio da pré-programação das partes que surgem com frequência. Mas essas habilidades e esse aprendizado não são reconhecidos: ao contrário, o termo "microtrabalho" adquiriu um significado diferente com o tempo, enfatizando a natureza mínima das tarefas, que se diz serem simples e curtas, e que não exigem qualificações. Aparentemente inexistente, a experiência adquirida com essa atividade nas plataformas não é, portanto, transferível de um serviço para outro, nem pode ser incluída em um currículo. Não é de se surpreender, portanto, que essa atividade pouco conhecida e desvalorizada não seja, até o momento, um trampolim para o sucesso no setor de tecnologia.

Portanto, estamos diante de uma ocupação que desempenha um papel essencial no desenvolvimento da IA, mas que, ao mesmo tempo, desvaloriza a inteligência humana que é empregada nela. Ao mesmo tempo, está aumentando a disparidade entre os grupos sociais mais privilegiados - os engenheiros e os profissionais de tecnologia de alto valor agregado que estão desenvolvendo a IA e que esperam colher os benefícios - e o lumpenproletariado da economia de plataforma, cujo tempo e habilidades estão sendo usados na IA, sem que seu papel ou contribuição sejam reconhecidos. Em escala global, há também uma lacuna cada vez maior entre os países produtores de IA, principalmente no Norte do mundo (bem como alguns países de renda média, como a China e a Índia), de onde vem a maior parte da demanda por microtrabalho, e os países do Sul, onde a oferta está aumentando.

Nossa contribuição, que documenta as condições do microtrabalho no Brasil e oferece retratos dos trabalhadores, é um passo no esforço mais amplo para superar o atual estado de invisibilização. Ele abre caminhos para pesquisas futuras, com o objetivo de caracterizar melhor essa nova forma de trabalho, mesmo que seja apenas em termos de terminologia, rastreando suas mudanças ao longo do tempo em relação à dinâmica da globalização e, idealmente, identificando alavancas para ação e transformação.

# Referências